\documentclass[twocolumn,showpacs,aps,prl]{revtex4}
\usepackage{graphicx}
\usepackage{dcolumn}
\usepackage{amsmath}
\usepackage{latexsym}

\begin {document}


\title {Conservative self-organized extremal model for wealth distribution}
\author
{
Abhijit Chakraborty$^1$, G. Mukherjee$^{1,2}$ and S. S. Manna$^1$
}
\affiliation
{
\begin {tabular}{c}
$^1$Satyendra Nath Bose National Centre for Basic Sciences,
Block-JD, Sector-III, Salt Lake, Kolkata-700098, India \\
$^2$Bidhan Chandra College, Asansol 713304, Dt. Burdwan, West Bengal, India \\
\end{tabular}
}
\begin{abstract}
      We present a detailed numerical analysis of the modified version of a conservative self-organized extremal model introduced 
   by Pianegonda et. al. for the distribution of wealth of the people in a society. Here the trading process has been modified by the 
   stochastic bipartite trading rule. More specifically in a trade one of the agents is necessarily the one with the 
   globally minimal value of wealth, the other one being selected randomly from the neighbors of the first agent. The pair 
   of agents then randomly re-shuffle their entire amount of wealth without saving. This model has most of the characteristics 
   similar to the self-organized critical Bak-Sneppen model of evolutionary dynamics. Numerical estimates of a number of critical 
   exponents indicate this model is likely to belong to a new universality class different from the well known models in the 
   literature. In addition the persistence time, which is the time interval between two successive updates of wealth of an agent
   has been observed to have a non-trivial power law distribution. An opposite version of the model has also been studied where 
   the agent with maximal wealth is selected instead of the one with minimal wealth, which however, exhibits similar behavior as 
   the Minimal Wealth model. 
\end{abstract}
\maketitle

\section {1. Introduction}

      Study of the probability distribution of wealth of the people in a society goes back to 1897 when Pareto observed
   empirically that the distribution is characterized by a power law tail. Probability that an individual
   member has wealth more than $w$ is given by $P(w) \sim w^{1-\alpha}$ with $\alpha = 5/2$ \cite {Pareto}. 
   This type of distribution is known as the Pareto distribution \cite {Wikipedia}. This observation reflects the inherent inequality 
   in the economic structure of the society. A large number of individuals are economically poor. In comparison the number
   of wealthier people is less and their number gradually decreases as their wealth increases. The cut-off
   of the distribution corresponds to with few very rich individuals. Consequently a sizable fraction of the 
   society's net wealth is infact possessed by a class of top rich people consisting only a few percent of 
   society's entire population. Since measuring wealth is difficult, in recent years distributions of 
   income and tax return amounts have been studied in different countries. For example tax return amount 
   distribution in US and Japan shows a log-normal distribution in the middle range followed by a power law for 
   high income people \cite {Souma}, UK data of income shows an exponential decay which is followed by a 
   power law in the high income range \cite {Dragulescu} and income data in Brazil for 2004 
   shows an almost Gaussian law for the low and middle income groups where as high salary groups are described
   approximately by power law \cite {Iglesias}.

      A good amount of research effort has been devoted in recent times to study the wealth distribution using 
   the ideas of Statistical Physics. In this description an individual member is called an agent. The microstate 
   specified by the precise description of wealth of every agent changes after each transaction. 
   Their wealth change due to interaction among themselves. This interaction is the mutual bipartite trade among 
   different pairs of agents. Thus the wealth distribution evolves due to such repeated interactions and finally 
   assumes a time independent form.

      In a simple model the total wealth of the entire society has been considered as strictly conserved. In a transaction the net
   amount of wealth of the pair of agents is randomly re-shuffled between them and therefore the pairwise interactions
   also maintain conservation of wealth \cite {DY,DY1}. Starting from an arbitrary initial distribution of wealth with a fixed
   average value the system attains a steady state. The steady state wealth distribution has been found to have an 
   exponentially decaying tail. Such distribution was also observed in \cite {Angle}. In a subsequent modification of the model saving propensity factor
   $\lambda$ has been introduced where each trader invests only $1-\lambda$ fraction of his wealth to the 
   bipartite trade \cite {CC}. With a fixed value of $\lambda$ the steady state wealth distribution becomes 
   very similar to the Gamma distribution \cite {Kunal,Patriarca}. This model was further extended by assigning 
   quenched saving propensities $\lambda_i$ specific to each individual agent \cite {CCM,Mohanty,BasuMohanty}. Here one gets a power 
   law tail with $\alpha \approx 2.0$ only when wealth distributions are averaged over different $\{\lambda_i\}$ sets \cite {CCM,Abhijit}.
   All these results exhibit economic inequality in the society. A review of all these models has been published in \cite {ArnabReview}.

\begin{figure*}[top]
\begin {center}
\includegraphics[width=15.0cm]{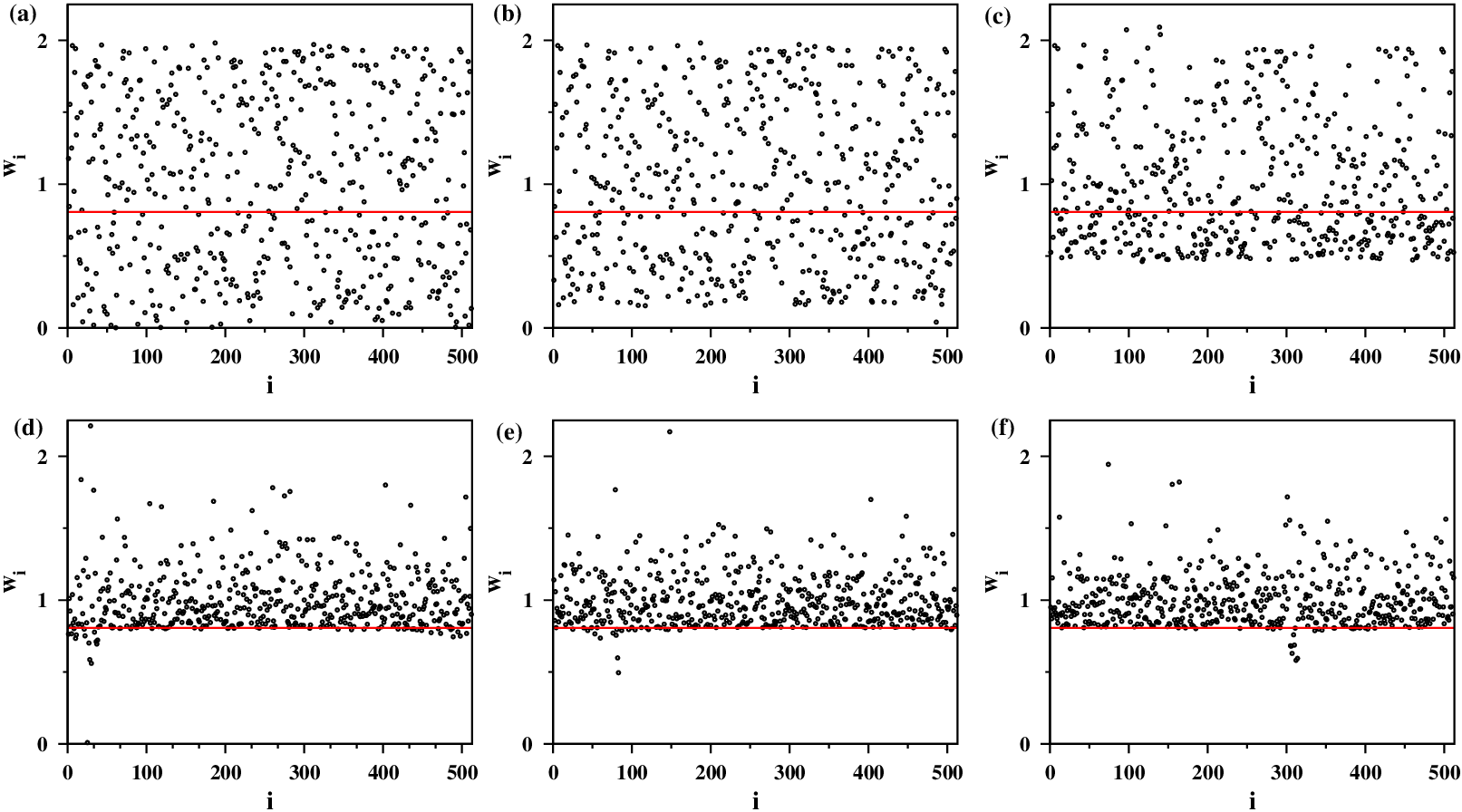}
\end {center}
\caption{(Color online) Time evolution of the wealth $w_i$ at different sites of an $1d$
lattice of size $N=512$. The series of snapshots are taken at times $t$ = (a) 0, (b) 500, (c) 1000,
(d) 500000, (e) 5000000 and (f) 15000000. The position of the poverty line at $w^{half}_c(512)$ = 0.8070 is shown by the
red line. With time the wealth values gradually move up beyond the poverty line and stays there in the
stationary state during further evolution. 
}
\end{figure*}

      Apart from these a different idea of extremal dynamics for the evolution of wealth distribution 
   has been studied by Pianegonda et. al. on an one dimensional ($1d$) lattice \cite {Pianegonda,Pianegonda1}. It is assumed 
   that always the poorest agent of the society initiates a trade since he feels the strongest urge to raise 
   its economic status. The trade is implemented by locating the poorest agent and refreshing his wealth 
   randomly. To ensure that the model remains strictly conserved, the amount of wealth gained by the 
   minimal site has been taken out equally from its two neighbors. Consequently this model allows an agent to possess a 
   negative wealth. In the stationary state of large systems $P(w)$ jumps from zero to a maximal value at a 
   critical value of $w_c$ and then it decays as $w$ increases as the Boltzmann-Gibbs exponential function.

      The Pianegonda model is very similar to the non-conservative self-organized extremal model for the 
   ecological evolution of interacting species introduced by Bak and Sneppen (BS) \cite {BS}. The phenomenon 
   of Self-organized Criticality (SOC) is the spontaneous emergence of fluctuations of all length and time 
   scales in a slowly driven system. This concept was first introduced to describe the formation of a sandpile 
   of a fixed shape \cite {Bak}. Later the idea of SOC has been applied to a large number of different physical 
   systems \cite {Bakbook}. In a stochastic version of the sandpile model grains are distributed to randomly 
   selected neighboring sites \cite {MannaSOC}. In the SOC models fluctuations are described in terms of avalanches 
   of activities and their size distributions assume power law decaying functions for large system sizes.
   In BS model an entire species is represented by a single fitness variable. A set of species is represented by 
   the nodes of a graph. Using the spirit of Darwinian principle in each mutation the fitness of the node 
   with globally minimal value is searched and is refreshed by a new random value. Effect of this 
   propagates to few neighboring nodes which are also refreshed. The system eventually reaches a stationary 
   state when the fitness distribution assumes a time independent step like form.

      An absorbing state phase transition in presence of a conserved continuous local field has been studied recently 
   \cite {Basu}. In this model a pair of sites is said to be active if at least one of them has energy more than a
   pre-assigned threshold. An active pair re-shuffles its net energy between them keeping the energy strictly conserved.
   Beyond a critical value of the threshold the number of active pairs fluctuates with time in the stationary state and 
   the time averaged density of active pairs has been considered as the order parameter for the problem. As the threshold 
   value is tuned a continuous phase transition is observed from an absorbing phase to an active phase \cite {Basu}.
   It was claimed that the critical exponents of this model are different from the Directed Percolation model.

      In a related model defined for the wealth distribution at least one of the agents in a bipartite trading is selected 
   from a subset of agents \cite {Ghosh}. This subset is formed by agents who have wealth less than certain upper cutoff,
   the other agent being selected randomly from the neighbors of the first agent. In a transaction the net wealth of the 
   pair of agents is randomly re-shuffled between them. The order parameter has been defined as the fraction of agents 
   having wealth below a certain threshold value and it is claimed that the system undergoes a continuous phase transition 
   at a critical value of the threshold wealth. A number of critical exponents have been measured to characterize the 
   transition and some of them are found to be close to corresponding exponents in the Manna model of Self-organized 
   Critically \cite {MannaSOC}.

      In this paper we present a detailed numerical analysis of the modified Pianegonda et. al. model whose trading rule 
   has been replaced by the stochastic re-shuffling of the net wealth of a pair of agents. We consider this model as one 
   of the few examples of non-dissipative SOC systems where the entire wealth of the society is strictly conserved, for 
   example the fixed energy sandpile \cite {Lubeck1}. Estimation of a number of critical exponents of the modified Pianegonda et. al. model suggests 
   that this model does not belong to the universality class of either BS model or Manna model. We believe that it belongs 
   to a new universality class perhaps because of strict conservation of wealth is maintained in its dynamical rules. The
   model studied by us in this paper and that in \cite {Ghosh} are essentially same but studied from two different approaches. 
   An approach which is very typical of a SOC system has been followed by us where the critical poverty line spontaneously evolves
   without any fine tuning. In contrast in \cite {Ghosh} the position of the poverty line is tuned by hand to arrive at the
   critical state.

\begin{figure}
\begin {center}
\includegraphics[width=7.0cm]{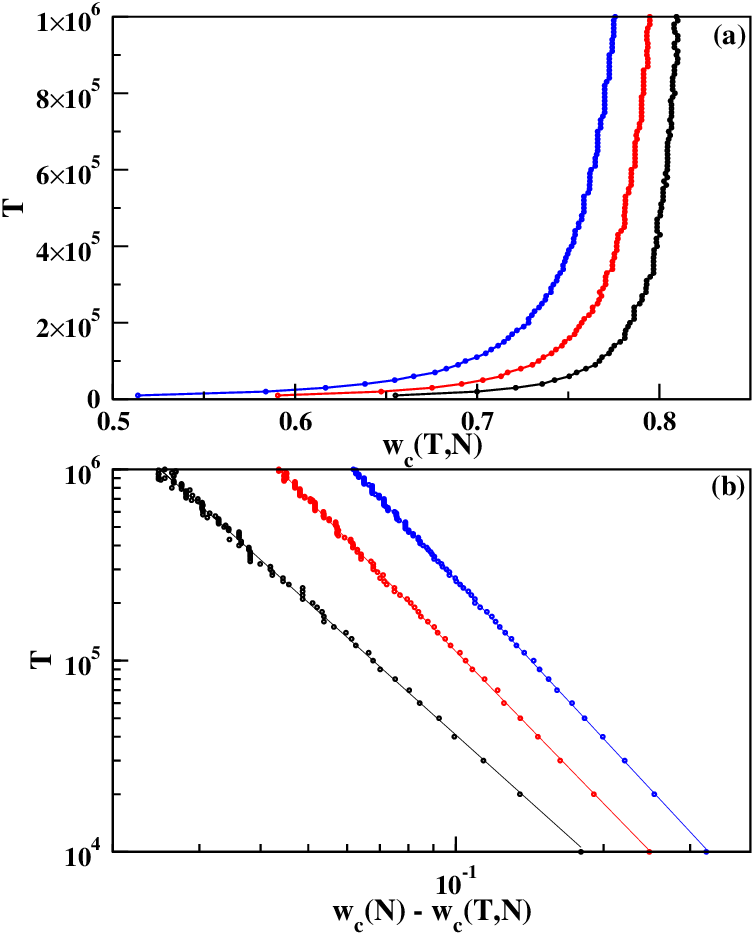}
\end {center}
\caption{(Color online) The system is relaxing from initial state to the stationary state for 
$N = 2^{10}$ (black), $2^{11}$ (red) and $2^{12}$ (blue).
(a) The relaxation time $T$ has been plotted with the corresponding position of the poverty line $w_c(T,N)$. 
(b) The same data has been plotted with deviation $w_c(N) - w_c(T,N)$ on a $\log - \log$ scale. The slopes are
-2.30, -2.64 and -2.76 respectively, which on extrapolation give the value of the dynamical exponent $z \approx 2.84$.
}
\end{figure}

      In section 2 we describe the Minimal Wealth model where the trader with minimal wealth initiates the trade. In subsection
   2.1 we discuss the relaxation dynamics of the system on its way to the stationary state. In subsection 2.2 the
   correlation that evolves in the system in the stationary state is studied. Wealth distribution in the stationary
   state is described in subsection 2.3. The statistics of the avalanche life-time distributions has been studied
   in section 2.4. This section ends with the study of persistence time distribution of individual agent's wealth
   in subsection 2.5. In section 3 we have described the Maximal Wealth model, which is opposite to the Minimal Wealth model, where 
   the trader with maximal wealth initiates the trade. Finally we conclude in section 4. 

\begin{figure}
\begin {center}
\includegraphics[width=7.0cm]{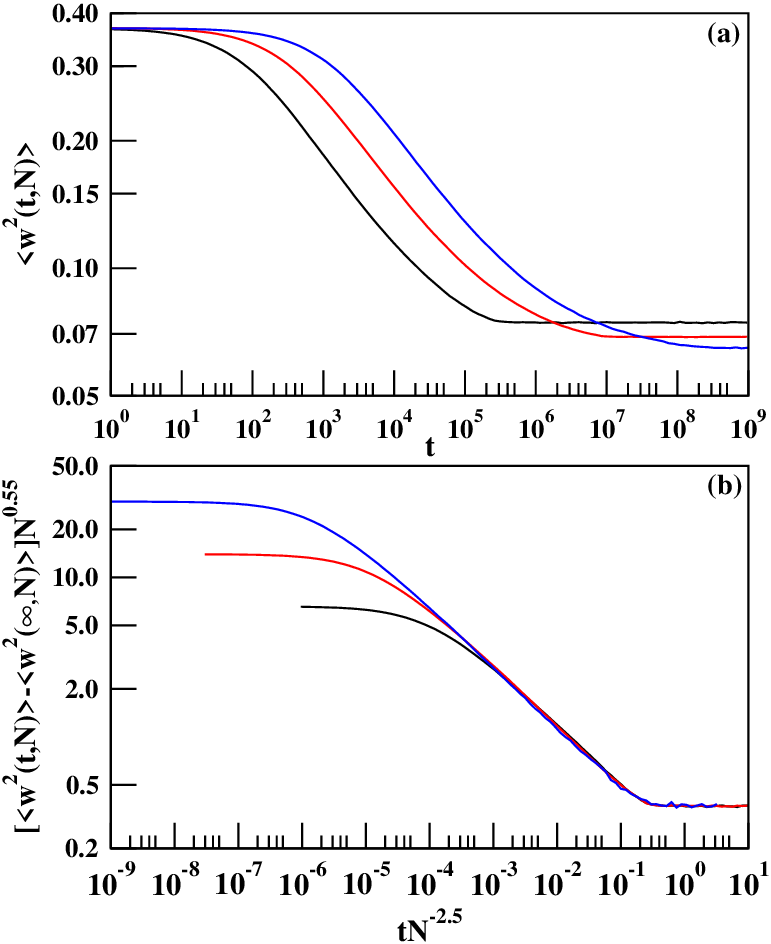}
\end {center}
\caption{(Color online) Estimating the relaxation times. (a) The average wealth square $\langle w^2(t,N) \rangle$ per agent has been 
plotted with time $t$ for $N = 2^{8}$ (black), $2^{10}$ (red) and $2^{12}$ (blue) and after a long time they approach the stationary state value
$\langle w^2(\infty,N) \rangle$. (b) The scaled deviation $[\langle w^2(t,N) \rangle - \langle w^2(\infty,N) \rangle]N^{0.55}$ 
has been plotted with the scaled variable $tN^{-2.5}$. The relaxation exponent $z = 2.70$.
}
\end{figure}

\section {2. The Minimal Wealth Model}

      In this paper we have considered a model with a conservative extremal dynamics for studying the evolution
   of wealth distribution in a society. In a bipartite transaction one agent is necessarily selected as the one
   with the globally minimal value of wealth $w_{min}$. The second agent is chosen randomly with uniform probability 
   from neighbors of the first agent. This neighbor list has been defined in different ways for different graphs. We 
   have studied this model on four different graphs, namely, (i) $1d$ regular lattice with periodic boundary condition (ii) two dimensional $(2d)$
   square lattice with periodic boundary condition (iii) the Barab\'asi - Albert (BA) scale-free graph \cite {Barabasi} and on an (iv) $N$-clique graph. 
   On every graph the nodes represent agents and the nearest neighbors of each node connected by direct links 
   constitute the neighbor list of every agent. We report elaborately the results of our model on an $1d$ 
   lattice and mention the key results of the same model studied on other graphs in tables.

      The dynamics starts with $N$ agents, each having an amount of wealth $w_i, \{i=1,N\}$ drawn from a uniform 
   distribution with the average $\langle w(N) \rangle$ =1 irrespective of the system size $N$. The discrete time 
   $t$ is the number of bipartite transactions. At an arbitrary time $t$ first the site $i=i_{min}$ is searched 
   out which has the minimal wealth $w_{min}$. The other agent $j$ is selected randomly with uniform probability 
   from the neighbors of $i_{min}$. Both agents invest their entire amount of wealth. Therefore 
   the total invested amount by both the traders is: $\delta_{ij}(t) = w_i(t)+w_j(t)$. This amount is then randomly 
   divided into two parts and received by them also randomly:
\begin {eqnarray}
& w_i(t+1)= \epsilon(t)\delta_{ij}(t) \hspace*{0.5 cm} w_j(t+1)= \bar {\epsilon}(t)\delta_{ij}(t).
\end {eqnarray}
   where $\epsilon(t)$ is a freshly generated random fraction and $\bar {\epsilon} = 1 - \epsilon$. These transactions 
   are repeated ad infinitum. After some relaxation time the system reaches a stationary 
   state when the wealth distribution assumes a time independent form.

\begin{figure}
\begin {center}
\includegraphics[width=8.5cm]{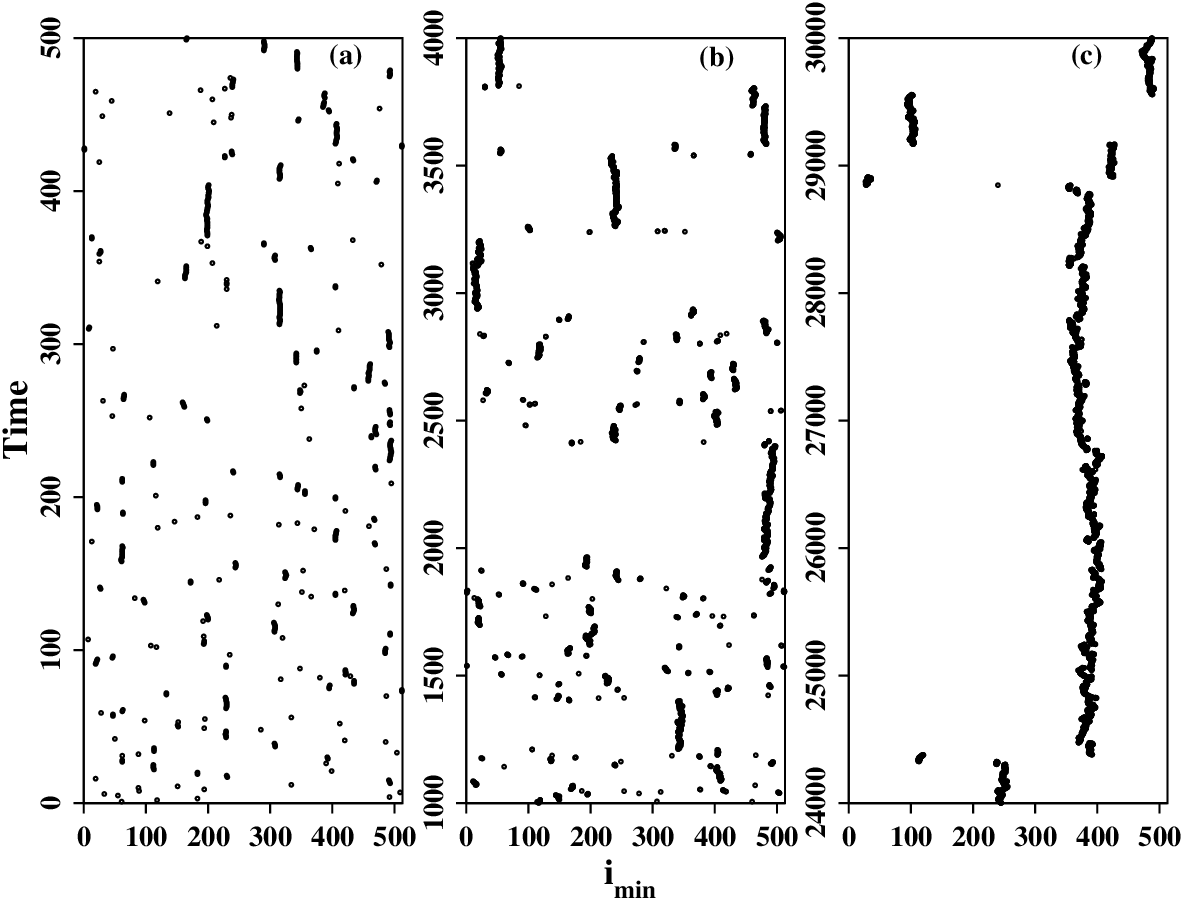}
\end {center}
\caption{An exhibition of the correlation that sets in the system. In a single run the lattice sites $i_{min}$ with 
globally minimal wealth in successive time steps are marked for a system of size $N$ = 512 in $1d$. The gradually increasing 
duration of correlation has been exhibited by time windows of increasing lengths: (a) 500, (b) 3000 and (c) 6000.
}
\end{figure}

      In $1d$ a linear chain of $N$ sites with periodic boundary condition has been considered where the neighbor list of 
   an arbitrary site $i$ consists of the two nearest neighboring sites at $i \pm 1$. Therefore 
   the second agent $j$ is selected randomly with equal probability from this list. If $w_{min}$ is very small then 
   the probability that it will be replaced by an even 
   smaller wealth after trade is also small. However this probability gradually increases as $w_{min}$ increases. As the sites with the 
   minimal values of wealth are systematically replaced, very soon all nodes with small $w$ values are replaced by larger 
   values of $w$ resulting a vacancy in the small $w$ region. This is explained pictorially in Fig. 1. On an $1d$ lattice 
   with $N$ = 512 we plot the lattice positions $i$ along the abscissa and the corresponding wealth $w_i$ along the ordinate. 
   As time increases a vacant region gradually forms for small values of $w$ for all sites. If on the average the wealth of none of the
   agent is below a certain threshold value $w$, it is called the `poverty line'. In Fig. 1 the poverty line 
   gradually moves up with time and finally settles at a critical value $w^{half}_c(512) = 0.8070$ at the stationary state. 
   This behavior is the same for all system sizes $N$ but with different values of $w_c(N)$. Unlike the model in 
   \cite {Ghosh} here the critical poverty line is spontaneously determined by the dynamical evolution of the system where no 
   fine tuning is necessary which is the distinctive signature of self-organization and we will see in the following 
   that the model exhibits critical behavior as well.

\begin{figure}
\begin {center}
\includegraphics[width=7.0cm]{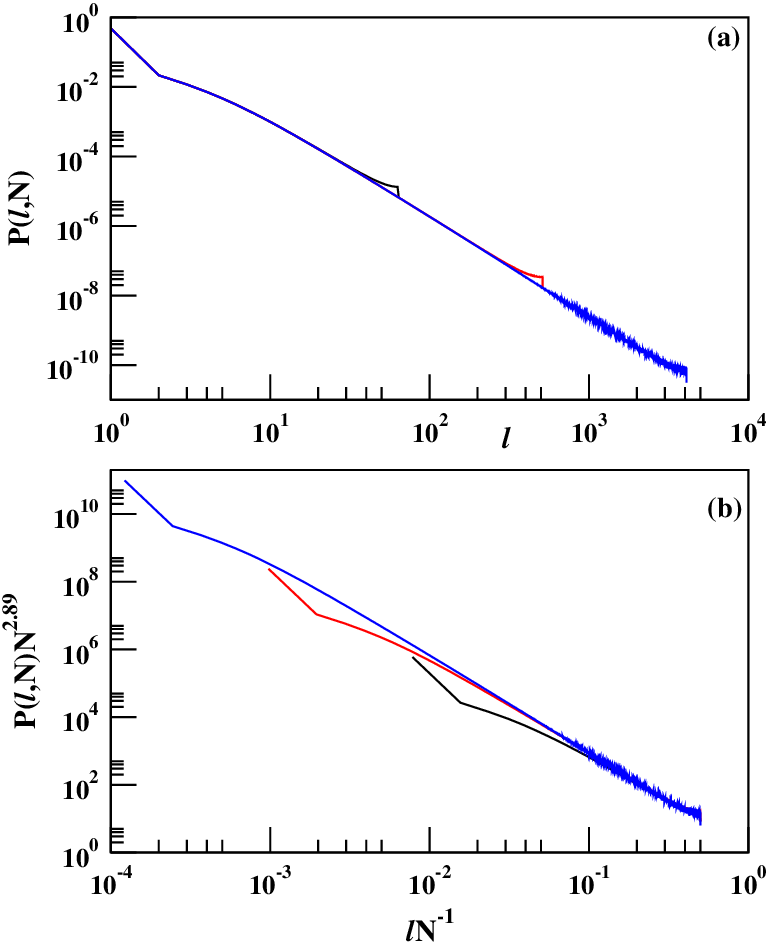}
\end {center}
\caption{(Color online) (a) The probability $P(\ell,N)$ that sites with minimal wealth at successive time steps are 
separated by a distance $\ell$ has been plotted with $\ell$ for the system sizes $N = 2^{7}$ (black), $2^{10}$ (red) and $2^{13}$ (blue). 
The slopes are -2.64, -2.81 and -2.89.(b) The finite-size scaling of $P(\ell,N)N^{\eta_{\pi}}$ vs. $lN^{-\zeta_{\pi}}$
with $\eta_{\pi}$ = 2.89 and $\zeta_{\pi}=1$ and therefore $\pi = \eta_{\pi} / \zeta_{\pi}=2.89(5)$.
}
\end{figure}

      To find the agent with minimal wealth a brute-force search takes cpu $\sim N$. A much faster algorithm to search 
   for the globally extremal (minimal or maximal) site was proposed by Grassberger \cite {Grassberger, Manna1} which stores 
   the data in a Hierarchical structure. This takes cpu $\sim \ln N$. We have used this method for $1d$, $2d$ and for $N$-clique 
   graphs. For BA graphs we used the brute-force algorithm.

\subsection {2.1 Relaxation to Stationary State}

      We first estimate the relaxation time required for the system to reach the stationary state. During this relaxation 
   period the wealth distribution gradually changes starting from its initial uniform distribution to its time independent 
   form in the stationary state. The relaxation time has been estimated as a function of deviation of the poverty line
   from its critical value in two ways. 

\begin{figure}
\begin {center}
\includegraphics[width=7.0cm]{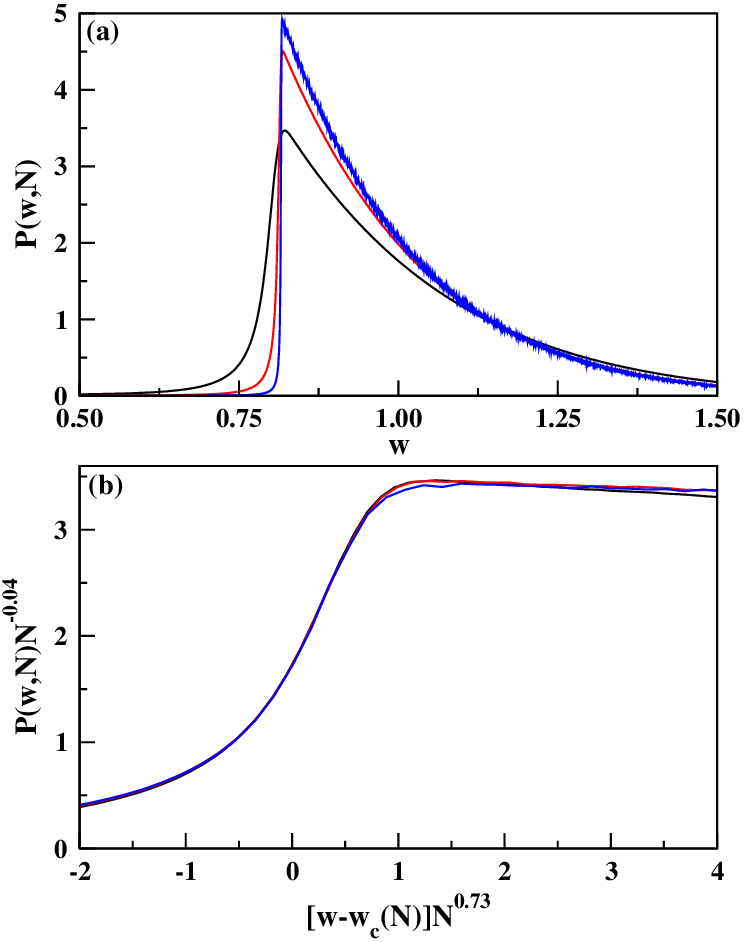}
\end {center}
\caption{(Color online) For Minimal Wealth model in $1d$. 
(a) The wealth distribution $P(w,N)$ in the stationary state for $N=2^7$ (black), $2^{10}$ (red) and $2^{13}$ (blue).
(b) A finite-size scaling using $P(w,N)N^{-0.04}$ vs. $[w-w_c(N)]N^{0.73}$ for system sizes $N=2^{11}$ (black), $2^{12}$ (red) and $2^{13}$ (blue).
}
\end{figure}
     
      For a given system size $N$ we have simulated our model up to $10^6$ time steps and calculated the wealth distribution 
   $P(T,w,N)$ at 100 time instants $T$ at the interval of $10^4$ steps. These distributions are calculated for a sample size of 
   $\sim 10^6$ independent runs. The value of the poverty line $w_c(T,N)$ at time $T$ is determined by the value of $w$
   where $P(T,w,N)$ is the maximum. This estimation is done for all values of $T$. In Fig. 2(a)
   we plot on a $lin - lin$ scale $T$ vs. $w_c(T,N)$ for $N = 2^{10}$, $2^{11}$ and $2^{12}$. We see that in each case the
   relaxation time $T$ diverges as $w_c(T,N)$ approaches its stationary state value $w_c(N)$. These data have been replotted
   in Fig. 2(b) using a $\log - \log$ scale as $T$ vs. $w_c(N) - w_c(T,N)$ where we have used $w_c(N)$ = 0.8242, 0.8375 and 
   0.8383 to obtain the best straight line plots. The slopes of these straight lines are 2.30, 2.64 and 2.76 respectively which are 
   then extrapolated with $N^{-1.503}$ to obtain the exponent $z$ as $T(N) \propto [w_c(N) - w_c(T,N)]^{-z}$ with $z \approx 2.84$.
   
      In a second method we calculated $\langle w^2(t,N) \rangle$ with time $t$ starting from its initial value when the distribution is 
   uniform. After a long time this quantity saturates to its stationary value $\langle w^2(\infty,N) \rangle$.  
   In Fig. 3(a) we show the plots of $\langle w^2(t,N) \rangle$ vs. $t$ on a $\log - \log$ scale.
   In Fig. 3(b) $[\langle w^2(t,N) \rangle - \langle w^2(\infty,N) \rangle]N^{0.55}$ has been plotted with the scaled 
   value of time $tN^{-2.5}$ using $\langle w^2(\infty,N) \rangle$ = 0.057, 0.0607 and 0.061 for $N = 2^{8}, 2^{10}$ and
   $2^{12}$ respectively. A nice data collapse has been obtained with the following scaling form
\begin {equation}
[\langle w^2(t,N) \rangle - \langle w^2(\infty,N) \rangle]N^{0.55} \sim {\cal F}_{z}(tN^{-2.5})
\end {equation}
   where the scaling function ${\cal F}_{z}(x)$ varies as $x^{-1/z}$ for small $x$. A direct measurement of the slope of the scaled
   plot gives an estimate of $1/z = 0.37$ which corresponds to $z = 2.70$. We conclude
   an average value of $z = 2.77(7)$.

\begin{figure}
\begin {center}
\includegraphics[width=7.0cm]{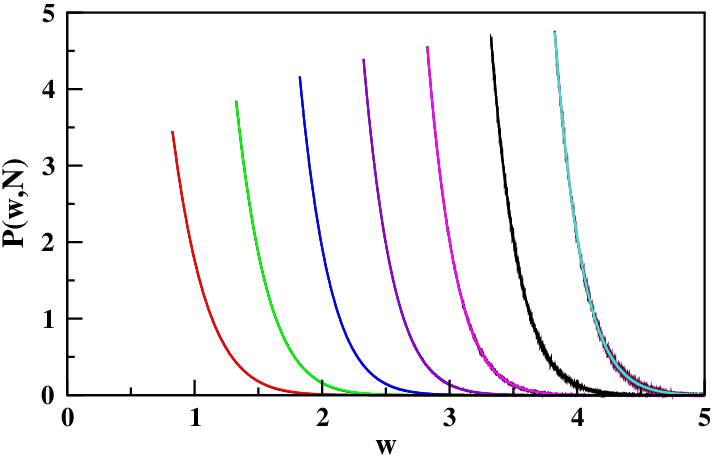}
\end {center}
\caption{(Color online) Gaussian fit of the wealth distribution $P(w,N)$ in the stationary state for seven 
different system sizes $N=2^7$ to $2^{13}$ in $1d$. The sequence starts with the distribution for $N=2^7$ at 
the extreme left and is shifted to the right by 0.2 when system size is multiplied by a factor of 2. 
}
\end{figure}

\subsection {2.2 Correlation in the Stationary State}

      Starting from an uncorrelated wealth distribution the system becomes more and more correlated as time
   passes. This is reflected in the fact that the probability of occurrence of the minimal sites close to each other
   in successive time steps gradually increases. At the early uncorrelated stage the position of the minimal site at
   the next time step is likely to be anywhere in the lattice. However as time increases, the poverty line moves up,
   consequently $w_{min}$ increases and the probability that the minimal site at the next time step occurring at the 
   same site or at the neighboring updated site also increases. This is shown in Fig. 4 using
   a $1d$ lattice of $N=512$. For a single run it is observed that the locations of $w_{min}$ are quite random (Fig. 
   4(a)). However as time evolves these positions gradually become more correlated (Figs. 4(b) and 4(c)). In general one can
   consider the successive jumps of $i_{min}$ positions constituting a L\'evy flight random walk \cite {Levy}. We see 
   below that indeed their flight lengths follow a power law distribution.

      The correlation in the stationary state is quantitatively measured by the probability distribution $P({\ell})$ 
   of the distance of separation ${\ell}$ between successive minimal sites using periodic boundary condition. This 
   distribution measured in the stationary state has been plotted in Fig. 5(a) for different system sizes $N = 2^7, 
   2^{10}$ and $2^{13}$. The value of $P({\ell})$ for ${\ell}$ = 0 and 1 are approximately 0.4575(1) and 0.4820(1) 
   and then it decreases as a power law $P({\ell}) \sim {\ell}^{-\pi}$ with increasing ${\ell}$. A direct measurement 
   of slope gives $\pi \approx 2.89$. Fig. 5(b) exhibits the finite-size scaling of the same data when the ${\ell}$ 
   and $P({\ell},N)$ axes are scaled as:
\begin {equation}
P(\ell,N)N^{\eta_{\pi}} \propto {\cal F_{\pi}}(\ell N^{-{\zeta_{\pi}}}).
\end {equation}
   where ${\cal F}_{\pi}(x)$ is a universal scaling function with the scaling exponents $\eta_{\pi}=2.89$ and $\zeta_{\pi} = 1$. 
   From this scaling analysis we get $\pi =\eta_{\pi}/\zeta_{\pi} = 2.89(5)$.

\begin{figure}
\begin {center}
\includegraphics[width=7.0cm]{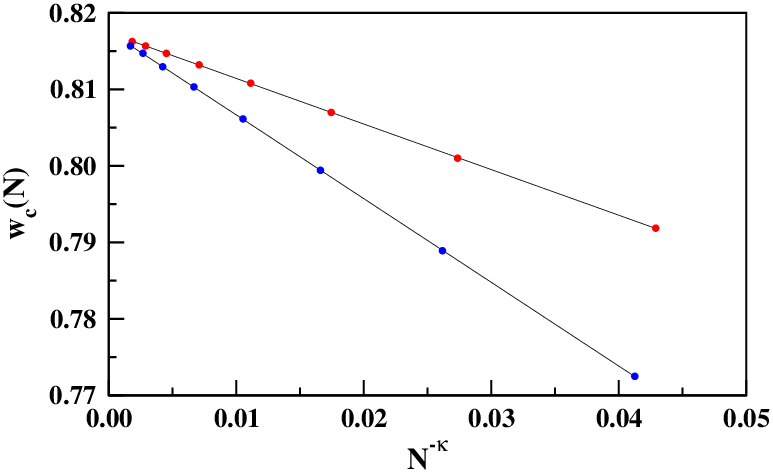}
\end {center}
\caption{(Color online) Estimation of the critical poverty line $w_c(\infty)$. The $w^{half}_c(N)$ are the upper plot (in red) and 
$w^{slope}_c(N)$ are the lower plot (in blue). When they are extrapolated with $N^{-\kappa}$ with $\kappa(half) = 0.649$ and $\kappa(slope) = 0.657$,
$w_c(\infty)$ is obtained as 0.8174 and 0.8176 respectively. We conclude $w_c(\infty)$ = 0.8175(2).
}
\end{figure}

      There exists a spatial correlation too in this model.
      A two point correlation function has been measured in the stationary state. The average correlation between 
   two sites situated at a distance of separation $x$ has been defined as:
\begin {equation}
{\cal C}(x) = \langle w(0)w(x) \rangle - \langle w \rangle^2
\end {equation}
   where $\langle w \rangle$ is always set equal to unity. We assume a power law decay for the correlation, i.e., ${\cal C}(x) \sim x^{-\chi}$
   for $x \to \infty$. For $1d$ a plot of (not shown) ${\cal C}(x)$ vs. $x$ on a $\log - \log$ scale indicates
   a power law for large $x$ values. However considerable variation of slopes exists for system sizes $N = 2^8, 2^{10}$ and $2^{12}$.
   The slopes are: -1.17, -1.29. -1.34 respectively which extrapolates to a value of $\chi = 1.5(1)$ in the limit of $N \to \infty$.
   Our estimate for $\chi$ in $2d$ is 2.2(2).

\begin{figure}
\begin {center}
\includegraphics[width=8.0cm]{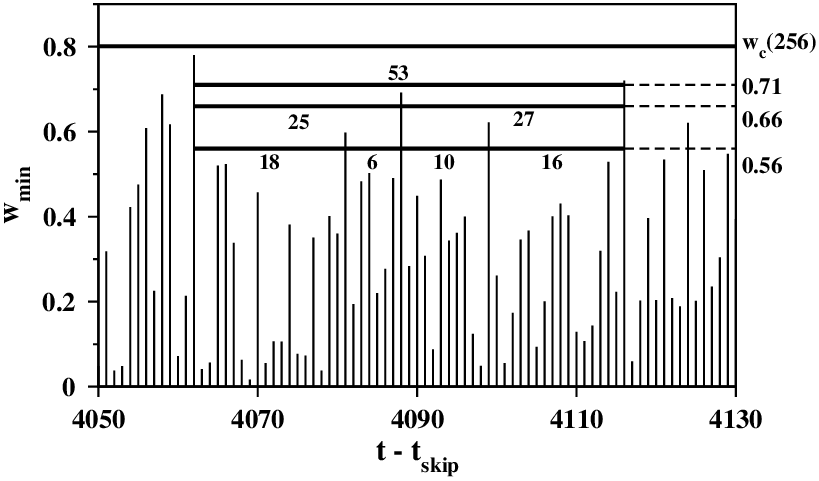}
\end {center}
\caption{A portion of the time series of minimal wealth values in successive time steps is shown for a $1d$ 
system with $N$ = 256 and $w_c(256)=w^{half}_c(256) = 0.8010$ in the stationary state. It shows that depending on the
value of $w_o$ an avalanche can be broken into a hierarchy of avalanches. For this run the system has been relaxed for the
initial $t_{skip} = 5 \times 10^8$ time steps. 
}
\end{figure}

\subsection {2.3 Wealth Distribution in the Stationary State}

      Next we estimated the probability density distribution $P(w,N)$ of wealth in the stationary state of
   the system of size $N$. The distribution grows very rapidly near $w_c(N)$ for all values of $N$ and then
   decays very fast. In Fig. 6(a) we show the plot of $P(w,N)$ vs. $w$ for $N = 2^7, 2^{10}$ and $2^{13}$. 
   All of them have similar variations but with increasing system size the curves gradually become sharper. 
   Therefore we tried a finite-size scaling analysis in Fig. 6(b) for the growing region and for $N = 2^{11}, 
   2^{12}$ and $2^{13}$. A nice data collapse is observed when axes are scaled and $P(w,N)N^{-0.04}$ has been 
   plotted with $[w-w_c(N)]N^{0.73}$. 

      The functional form of the decay of the probability distribution $P(w,N)$ has been studied right after the maximal jump
   at $w_c(N)$. This part fits very well with the Gaussian form: 
\begin {equation}
P(w,N) = \frac {A(N)}{\sqrt{2\pi}\sigma(N)}\exp [-\frac {(w-\mu(N))^2}{2\sigma^2(N)}]
\end {equation}
   In Fig. 7 we showed $P(w,N)$ vs. $w$ on a $lin - lin$ scale for seven different system sizes: $N = 2^7, 2^8, ... , 2^{13}$.
   In each case the fitting curve is indistinguishable from the data. We observe a systematic variation of $A(N)$, $\mu(N)$ and
   $\sigma(N)$ with system size $N$. For example $A(N) \approx 511.5 - 1828N^{-0.294}$, $\mu(N) \approx -1.10 + 62.5N^{-1.215}$ and
   $\sigma(N) \approx 0.64 + 0.429N^{-0.436}$.

      The precise value of $w_c(\infty)$ is calculated by extrapolating $w_c(N)$ values which are calculated
   using the following two methods. We have seen in Fig. 6(a) that the probability distribution of $P(w,N)$ 
   becomes increasingly steeper with increasing $N$. For a certain size $N$ we defined $w^{half}_c(N)$ as the 
   value of $w$ for which $P(w,N)$ is half of its maximum value. In a second method the $w_c(N)$ has been 
   calculated in the following way. A pair of successive points on the $P(w,N)$ vs. $w$ curve which has the 
   largest slope is found out. A straight line joining these two points is then extrapolated to meet the $w$ 
   axis at $w^{slope}_c(N)$. The pair of values of $w^{half}_c(N)$ and $w^{slope}_c(N)$ for eight $N$ values 
   $2^7$ to $2^{14}$ are then extrapolated with $N^{-\kappa}$. A least square fit of straight line has been 
   done for trial values of $\kappa$ starting from 0.20 to 1.20 at an interval of 0.001 and the errors have 
   been calculated. The errors are minimal for $\kappa(half) = 0.649$ and $\kappa(slope) = 0.657$. Using these 
   two values of the exponent $\kappa$ we extrapolate $w_c(N)$ values with $N^{-\kappa}$ in Fig. 8 to meet 
   the $w_c(N)$ axis at 0.8174 and 0.8176 respectively. We conclude a value for $w_c(\infty)$ for $1d$ model 
   as 0.8175(2).

\begin{figure}
\begin {center}
\includegraphics[width=7.0cm]{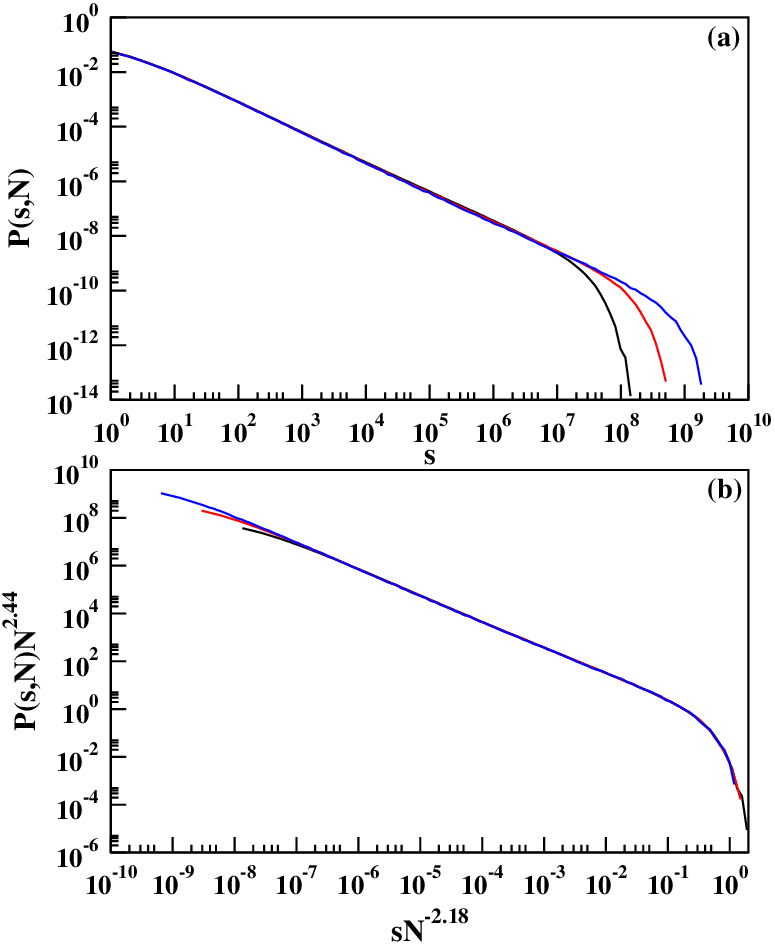}
\end {center}
\caption{(Color online) For $1d$ Minimal Wealth model. 
(a) The avalanche size distribution for $N=2^{12}$ (black), $2^{13}$ (red) and $2^{14}$ (blue).
(b) A finite-size scaling of the data in (a) with scaling exponents $\eta_{\tau}=2.44$ and $\zeta_{\tau}=2.18$
giving the avalanche size exponent $\tau=\eta_{\tau}/\zeta_{\tau} \approx 1.12(1)$.
}
\end{figure}

\subsection {2.4 Avalanche Size Distribution}

      In the stationary state successive values of minimal wealth $w_{min}$ fluctuates with time. If a certain reference wealth is
   fixed by hand at $w=w_o$ then the successive $w_{min}$ appear below and above $w_o$ line. One defines a $w_o$-avalanche as the 
   sequence of successive bipartite trades whose $w_{min}$ values are smaller than $w_o$. The size $s$ of the avalanche 
   measures the duration of the avalanche i.e., at times $t$ and $t+s+1$ the $w_{min} > w_o$, whereas at every time step from $t+1$ 
   to $t+s$ the $w_{min} < w_o$. When $w_o$ is set equal to $w_c(N)$ it is called a critical avalanche. This is explained in Fig. 9
   where part of the $w_{min}$ time series for $N = 256$ and $w_c(256)=w^{half}_c(256)$ = 0.8010 is displayed discarding the initial 
   $t_{skip}=5 \times 10^8$ time steps. For $w_o=0.71$ an avalanche of size 53 breaks into two avalanches
   of sizes 25 and and 27 when $w_o$ is reduced to 0.66. On further reduction of $w_o$ to 0.56 these two avalanche break
   into even smaller avalanches of sizes 18, 6 and 10, 16 respectively. Thus any avalanche can be splitted into a hierarchy 
   of smaller avalanches if $w_o$ value is lowered \cite {Paczuski}. On the other hand if $w_o$ is raised the average avalanche size
   increases and becomes infinite at certain value of $w_o$. 

\begin{figure}
\begin {center}
\includegraphics[width=7.0cm]{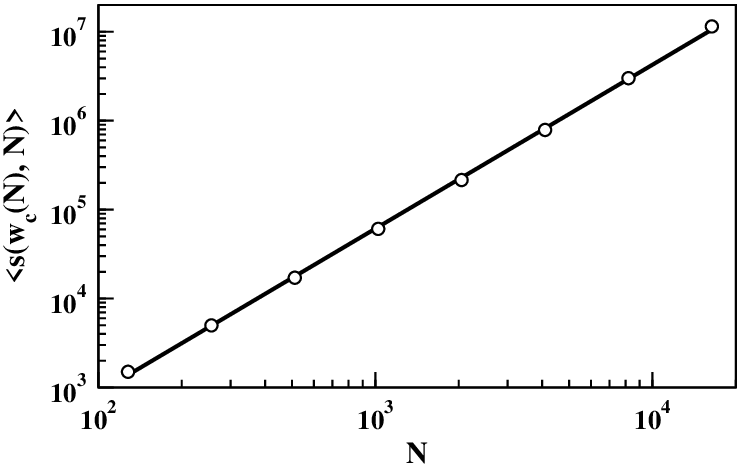}
\end {center}
\caption{The average size of the avalanches $\langle s(w_c(N),N) \rangle$ right at the poverty line 
$w_c(N)$ has been plotted with system size $N$ on a $\log-\log$ scale. The slope of this curve gives the exponent
$\beta = 1.92(2)$.
}
\end{figure}

      At the critical point the distribution of the avalanche life-times has a power law tail in the limit of $N \to \infty$: 
   $P(s,\infty) \sim s^{-\tau}$. In the stationary state we used $w_o = w^{half}_c(N)$ and measured life-times of a large 
   number of avalanches for different system sizes and plot the probability distributions $P(s,N)$
   vs. $s$ using a $\log - \log$ scale in Fig. 10(a). Each curve has a straight portion in the intermediate
   regime of the avalanche sizes and this regime becomes gradually larger on increasing $N$. The direct measurement of
   slopes in the scaling region gives $\tau(N)$ = 1.086, 1.091 and 1.096 for $N = 2^{12}, 2^{13}$ and $2^{14}$ respectively.
   A finite-size scaling is very much suitable with the following scaling form:
\begin {equation}
P(s,N)N^{\eta_{\tau}} \propto {\cal F}_{\tau}(sN^{-\zeta_{\tau}})
\end {equation}
   where the scaling function ${\cal F}_{\tau}(x) \sim x^{-\tau} $ in the limit of $x \to 0$ and
   ${\cal F}_{\tau}(x)$ approaches zero very fast for $x >> 1$. The exponents $\eta_{\tau}$ and $\zeta_{\tau}$ fully
   characterize the scaling of ${P(s,N)}$ in this case. An immediate way to check the validity
   of this equation is to attempt a data collapse by plotting $P(s,N)N^{\eta_{\tau}}$ vs. $s/N^{\zeta_{\tau}}$
   with trial values of the scaling exponents. For $1d$ the values for obtaining the best data collapse are found
   to be $\eta_{\tau}=2.44$ and $\zeta_{\tau}=2.18$ (Fig. 10(b)). The life-time distribution exponent for $1d$ is therefore $\tau=\eta_{\tau}/\zeta_{\tau} 
   \approx 1.12(1)$.

      Next we calculated the average value of avalanche life-times $\langle s(w_c,N) \rangle$ right
   at the critical poverty line. In Fig. 11 we plot this quantity with system size $N$ on a $\log - \log$
   scale. The plot fits excellent to a straight line and its slope gives the value of the exponent 
   $\beta \approx 1.92(2)$ in: $\langle s(w_c,N) \rangle \sim N^{\beta}$. Assuming the distribution 
   ${P(s,N)}$ of avalanche sizes holds good up to a cut-off $s_{max} \sim N^{\zeta_{\tau}}$ one gets a scaling relation 
   $\beta = \zeta_{\tau} (2-\tau)$ and our estimates of $\beta=1.92$, $\zeta_{\tau}=2.18$ and $\tau=1.12$ satisfy 
   this relation very closely.

\begin{figure}
\begin {center}
\includegraphics[width=7.0cm]{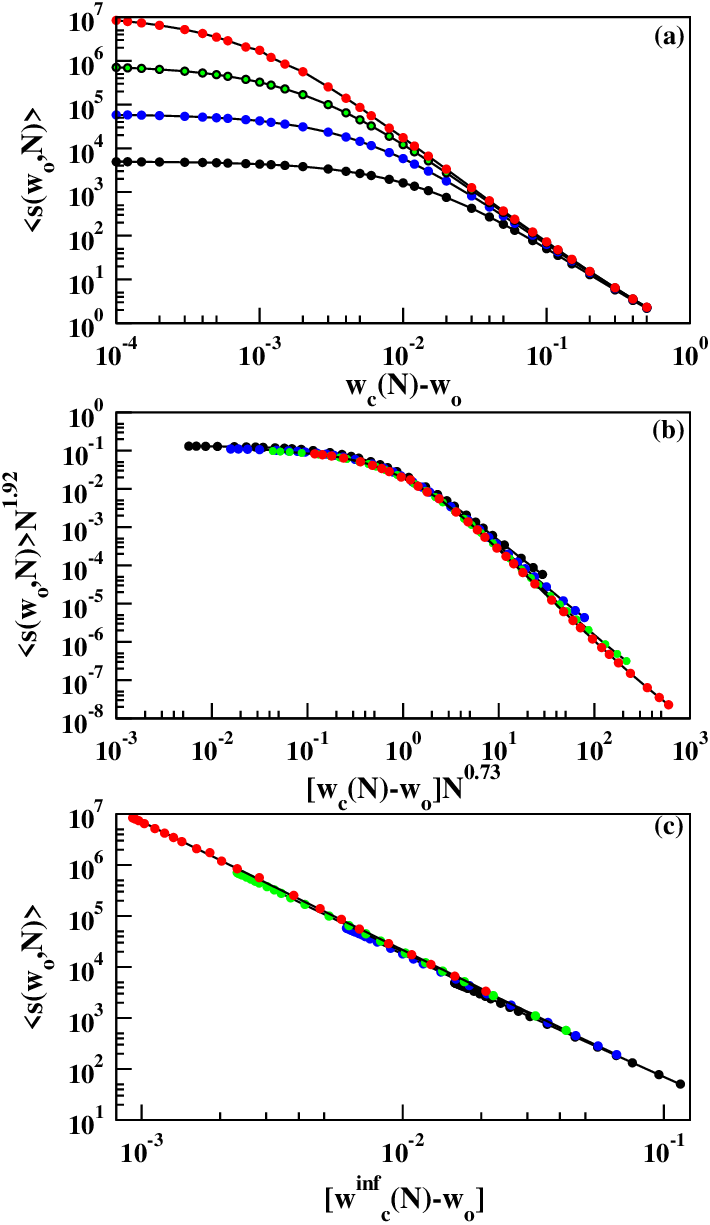}
\end {center}
\caption{(Color online) (a) The average value of the avalanche size $\langle s(w_o,N) \rangle$ has been plotted with
deviation $w_c(N)-w_o$ from the poverty line $w_c(N)$. The system sizes used are $N = 2^8$ (black), $2^{10}$ (blue), $2^{12}$ (green) and $2^{14}$ (red).
The value of $\gamma$ obtained by extrapolation of slopes is 2.67.
(b) Finite size scaling analysis of the data in (a) is shown. The scaling exponents $\eta_{\gamma}=\beta=1.92$ and $\zeta_{\gamma}=0.73$ gives
$\gamma=\eta_{\gamma}/\zeta_{\gamma} \approx 2.63$.
(c) Plot of the data in (a) but with $w_c(N)=w_c^{inf}(N)$. The slopes for four different system sizes in (a) on
extrapolation gives $\gamma = 2.66$. We conclude $\gamma = 2.65(5)$.
}
\end{figure}

      The size of the $w_o$-avalanches are smaller when $w_o < w_c(N)$ and we have studied how the average avalanche 
   size grows as the deviation $(w_c(N)-w_o)$ decreases. Similar to the BS model we assume $\langle s(w_o) \rangle \sim 
   [w_c - w_o]^{-\gamma}$ for $N \to \infty$. We measured the average size $\langle s(w_o,N) \rangle$ of the $w_o$-avalanches 
   for different system sizes $N$ and plotted them with $w_c(N)-w_o$ in Fig. 12(a) with $w_c(N) = w^{half}_c(N)$. For all 
   plots on $\log - \log$ scale the curves 
   are horizontal as deviation $w_c(N)-w$ is very small. However as the deviation increases the curves become straight 
   with negative slopes -1.98, -2.15, -2.28 and -2.38 for for $N = 2^8, 2^{10}, 2^{12}$ and $2^{14}$ 
   respectively. These values when extrapolated with $N^{-0.208}$ give $\gamma = 2.67$ for $N \to \infty$. Again a finite-size
   scaling has been possible as shown in Fig. 12(b):
\begin {equation}
\langle s(w_o,N) \rangle N^{-\eta_{\gamma}} \propto {\cal F}_{\gamma}([w_c(N)-w_o]N^{\zeta_{\gamma}})
\end {equation}
   where ${\cal F}_{\gamma}(x)$ is another scaling function. From this data collapse the scaling exponents
   $\eta_{\gamma}=\beta=1.92$ and $\zeta_{\gamma}=0.73$ with $\gamma = \eta_{\gamma} / \zeta_{\gamma} \approx 2.63$ is obtained.

\begin{figure}
\begin {center}
\includegraphics[width=7.0cm]{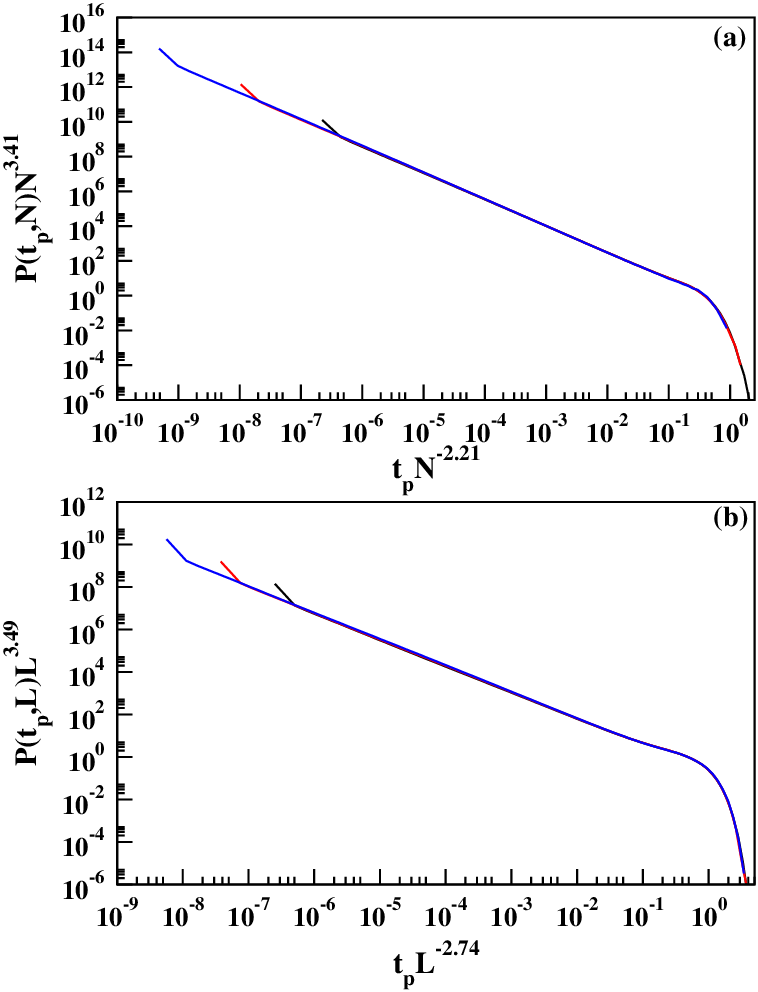}
\end {center}
\caption{(Color online) 
The probability distribution $P(t_p,N)$ of persistence times $t_p$ at the stationary state.
(a) The finite-size scaling of the distribution in $1d$ for $N = 2^{10}$ (black), $2^{12}$ (red) and $2^{14}$ (blue). From the scaling exponents
    $\eta_{\theta} = 3.41$ and $\zeta_{\theta} = 2.21$ the persistence exponent $\theta = \eta_{\theta} / \zeta_{\theta} \approx 1.543$ is obtained.
(b) The finite-size scaling of the distribution in $2d$ for $L = 2^{8}$ (black), $2^{9}$ (red) and $2^{10}$ (blue). From the scaling exponents
    $\eta_{\theta} = 3.49$ and $\zeta_{\theta} = 2.74$ the persistence exponent $\theta = \eta_{\theta} / \zeta_{\theta} \approx 1.274$ is obtained.
}
\end{figure}

      For every system size $N$ there is a value of $w_o=w^{inf}_c(N)$ so that when $w_o$ is raised to this value
   the avalanche size becomes infinite. This implies that if we plot the data in Fig. 12(a) with respect to 
   $w^{inf}_c(N)-w_o$ then we should be able to see the divergence of average avalanche size instead of saturation
   of the avalanche sizes. We plot this in Fig. 12(c) using $w^{inf}_c(N)$ = 0.8167, 0.8169, 0.8170, 0.8172
   for $N = 2^8$, $2^{10}$, $2^{12}$ and $2^{14}$ respectively again on a $\log - \log$ scale. Each curve is a
   straight line but with different slopes: -2.31, -2.43, -2.51 and -2.56 respectively. When these slopes are extrapolated
   with $N^{-0.31}$ the extrapolated value for $N \to \infty$ is -2.66. Our conclusion for the value of the exponent $\gamma = 2.65(5)$.

\begin{figure}
\begin {center}
\includegraphics[width=7.0cm]{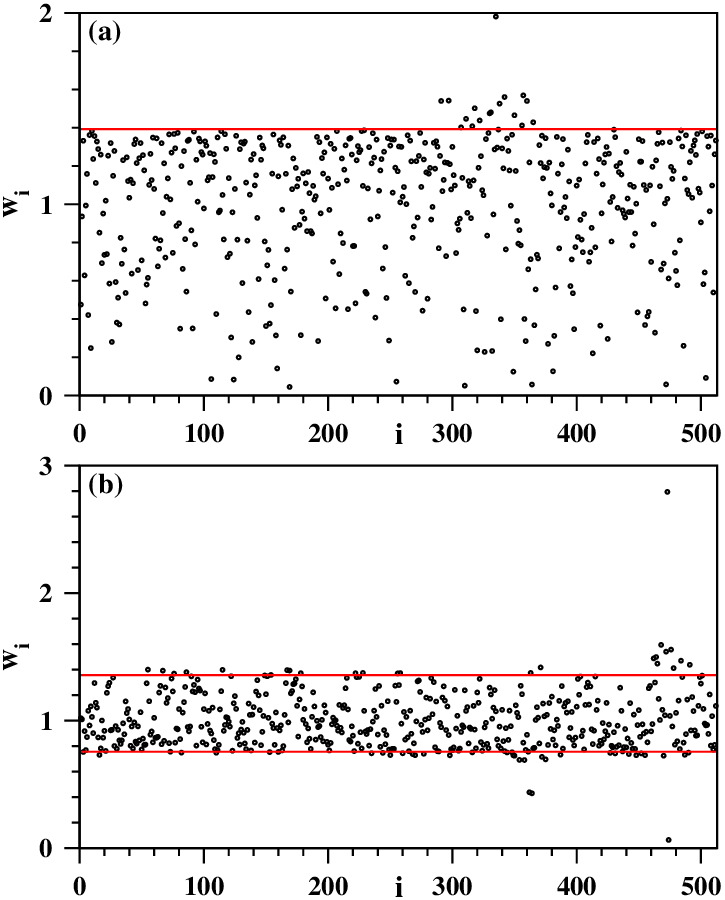}
\end {center}
\caption{(Color online) Values of wealth $w_i$ of different agents in the stationary state are plotted with their positions
$i$ along a $1d$ lattice for the 
(a) Maximal Wealth model, the red line is at $w^{half}_c(512)=1.3924$. 
(b) mixture of the Minimal Wealth and Maximal Wealth models with probability $p=1/2$, the red lines correspond to peaks at $w_c(512)=0.7559$ and at 1.3574.
}
\end{figure}

\subsection {2.5 Persistence of Wealth in the Stationary State}

      The time interval between two successive updates of wealth of an agent is known as the persistence time $t_p$. Different 
   agents have to wait for different amounts of times in general. More specifically agents having
   small amount of wealth have to wait for very little times. On the other hand potentially rich agents have to wait long
   enough times. In the stationary state we measure the persistence times for all sites of the lattice and use this data 
   to plot their probability distribution. More precisely we set a clock to each site. Whenever there is a change of wealth 
   at this site the time is noted and the clock time is reset to zero. At the stationary state we collect a large number of persistence 
   time data and use these data to calculate the persistence time distribution.

      We assume a power law variation of the persistence time distribution as $P(t_p) \sim t_p^{-\theta}$ in the limit
   of $N \to \infty$. For finite size systems the distributions $P(t_p,N)$ vs. $t_p$ are plotted on a $\log - \log$ scale 
   (not shown) and the direct measurement of slopes give the $\theta(N)$ values for finite size systems. These values are 
   extrapolated as $N^{-0.494}$ to obtain $\theta$ = 1.539 for $N \to \infty$ in $1d$. In a similar analysis for a $2d$ square lattice of size $L$ using 
   an extrapolation with respect to $L^{-0.565}$ we get $\theta$ = 1.25 for $L \to \infty$. 

      Persistence exponents are also 
   obtained by the finite-size scaling analysis. In Fig. 13(a) we show the scaling plot of $P(t_p,N)N^{\eta_{\theta}}$ vs. 
   $t_pN^{-{\zeta_{\theta}}}$ with $\eta_{\theta}=3.41$ and $\zeta_{\theta}=2.21$. This gives $\theta = \eta_{\theta} / 
   \zeta_{\theta} = 1.543$ in $1d$. Similar scaling analysis in terms of the system size $L$ in $2d$ square lattice has been 
   performed with $\eta_{\theta}=3.49$ and 
   $\zeta_{\theta}=2.74$ which gives $\theta$ = 1.274 (Fig. 13(b)). Averaging $\theta$ values obtained from direct measurement 
   and scaling analysis we conclude $\theta = 1.541(10)$ in $1d$ and $\theta = 1.262(10)$ in $2d$.

\begin{table*}[top]
\begin {center}
\begin{tabular}{|c|r|r|r|r|r|r|r|r|r|r|} \hline
        & \multicolumn{4}{c|}{Minimal Wealth model}& \multicolumn{2}{c|}{BS model} & \multicolumn{2}{c|}{Manna model} 
\\ \cline{2-5} \cline{6-7} \cline{8-9} 
        & $1d$       & $2d$       & BA graph    & $N$-clique & $1d$     & $2d$         & $1d$  & $2d$  \\ \hline
$w_c   $& 0.8175(2)  & 0.6887(2)  & 0.6444(2)   & 0.6076(2)  & 0.66702(8) \cite {Grassberger} & 0.328855(4) \cite {Paczuski}& 0.89199(5) \cite {Lubeck} 
& 0.68333(3) \cite {Lubeck} \\
$\tau  $& 1.12(1)    & 1.29(1)    & 1.50(1)     & 1.50(1)    & 1.073(3) \cite {Grassberger}  &1.245(10) \cite {Paczuski}  & 1.112(6) \cite {Huynh} & 1.273(2) \cite {Huynh} \\
$\gamma$& 2.65(5)    & 1.58(5)    & 1.02(5)     & 1.00(5)    & 2.70(1) \cite {Paczuski}   &1.70(1) \cite {Paczuski}    &       &             \\
$\pi   $& 2.89(5)    & 3.94(5)    & -           & -          & 3.23(2) \cite {Paczuski}   &            &       &           \\ 
$z     $& 2.77(7)    &            &             &            &            &            &       &       \\ 
$\beta $& 1.92(2)    & 0.95(2)    & 0.52(2)     & 0.50(2)    &            &            & 2     & 2                \\ 
$\theta$& 1.541(10)  & 1.262(10)  & 1.00(1)     & 1.00(1)    &            &            &       &                 \\ \hline
\end{tabular}
\caption{Values of different exponents of Minimal Wealth model are compared with those of existing models in the literature.}
\end {center}
\end {table*}

\section {3. The Maximal Wealth model}

   Next we studied the Maximal Wealth model where one agent is necessarily the agent with maximal wealth.
The other agent being selected randomly with uniform probability from the neighbors of the first agent.
Random re-shuffling of wealth takes place in the same way as in the Minimal Wealth model.

\begin{figure}
\begin {center}
\includegraphics[width=7.0cm]{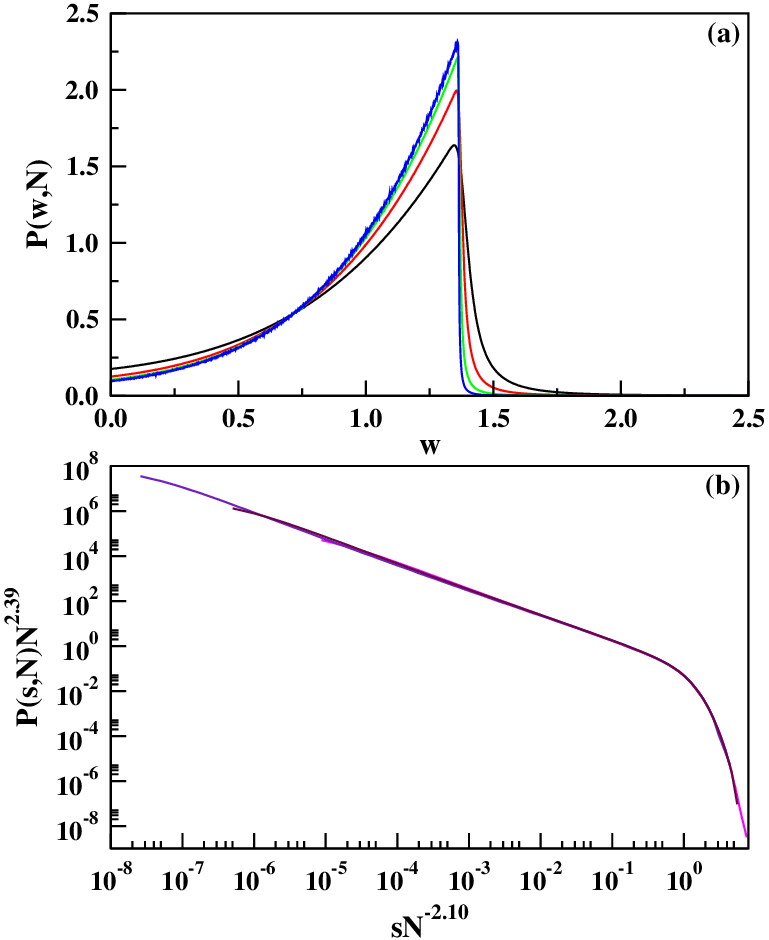}
\end {center}
\caption{(Color online) (a) The wealth distribution $P(w,N)$ vs. $w$ for the Maximal Wealth model for $N$ = $2^8$ (black), $2^{10}$ (red), $2^{12}$ (green) and
for $2^{14}$ (blue). (b) Finite-size scaling of the avalanche size distribution $P(s,N)$ at the critical threshold $w_c(N)$ for $N$ = $2^8$ (black), 
$2^{10}$ (red) and $2^{12}$ (blue). The scaling exponents are $\eta_{\tau} = 2.39$ and $\zeta_{\tau} = 2.10$ which gives the exponent 
$\tau = \eta_{\tau} / \zeta_{\tau} \approx 1.14(1)$.
}
\end{figure}

\begin{figure}
\begin {center}
\includegraphics[width=7.0cm]{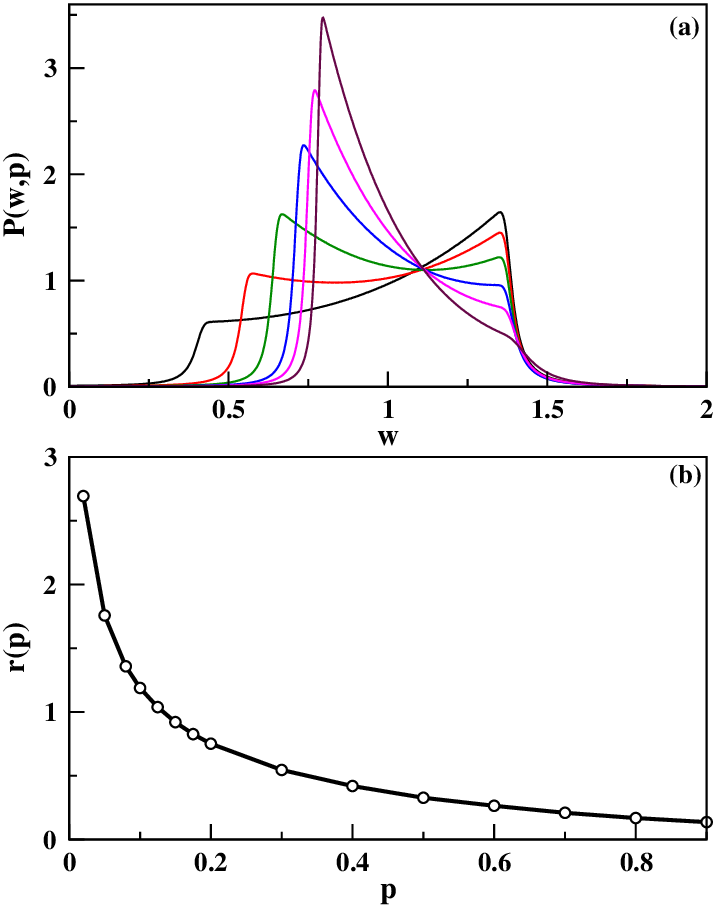}
\end {center}
\caption{(Color online) (a) Mixture of Minimal Wealth and Maximal Wealth models with probabilities $p$ and $1-p$ respectively for
$N=2^9$ and for $p$ = 0.02 (black), 0.08 (red), 0.2 (green), 0.4 (blue), 0.6 (magenta) and 0.9 (maroon). (b) Variation of the ratio $r(p)$ of heights of the right peak and the left peak
with probability $p$.
}
\end{figure}

   In Fig.14(a) we plot again for the Maximal Wealth model the values of wealth $w_i$ of different agents 
at a certain instant of time in the stationary state with their positions $i$ along an $1d$ lattice of size 
$N=512$. In contrast to the similar plot of the Minimal Wealth model in Fig. 1 here an upper cut-off 
for the wealth has been visible at $w^{half}_c(512) = 1.3924$.
 
   In this case the stationary state wealth distribution $P(w,N)$ takes an opposite shape (Fig. 15(a)). 
A critical wealth $w_c(N)$ exists here as well. $P(w,N)$ takes a Gaussian form elevated
by a constant term $c(N)$ for $w < w_c(N)$, whereas for $w > w_c(N)$ it sharply decreases to zero. The parameters
of the Gaussian function (Eqn. (4)) are different for different $N$ and they vary very systematically
with $N$ as: $A(N) \approx 30.48 - 133.35N^{-0.361}$, $\mu(N) \approx 3.74 -6.66N^{-0.85}$ and
$\sigma(N) \approx 1.047 + 0.572N^{-0.321}$ and the constant $c(N) \approx 0.031 + 4.265N^{-0.73}$.

   The critical value of wealth $w_c(\infty)$ in the asymptotic limit of system sizes has been estimated
by the same method as used for the Minimal Wealth model. The $w^{half}_c(N)$ and $w^{slope}_c(N)$ values have been
calculated for $N = 2^8$, $2^{10}$, $2^{12}$ and $2^{14}$, extrapolated with $N^{-0.586}$ and $N^{-0.620}$
and the asymptotic values are 1.3610 and 1.3608 respectively. We conclude $w_c(\infty) = 1.3609(2)$ for $1d$.
A similar analysis gives $w_c(\infty) = 1.7076(2)$ for $2d$ square lattice, 1.8895(2) for the BA graph and 1.9998(2) for the $N$-clique.

   It may appear that the Minimal Wealth and Maximal Wealth models should be symmetric about the average
wealth per agent which we have set at $\langle w \rangle =1$. We have seen above that this indeed not the
case since $w_c$ values are 0.8175 and 1.3609 for the Minimal Wealth and Maximal Wealth models respectively.
The symmetry between these two models are broken by the presence of a rigid wall at $w=0$ which means
that negative value of wealth of an agent is not allowed. 

   The avalanche size distributions have been studied as well. A finite-size scaling of these 
distributions has been done and are plotted in Fig. 15(b) using $\log - \log$ scale as before for 
$N = 2^{8}, 2^{10}$ and $2^{12}$. The scaling exponents are $\eta_{\tau} = 2.39$ and $\zeta_{\tau} = 2.10$ respectively
giving the value of the avalanche size exponent $\tau = \eta_{\tau} / \zeta_{\tau} \approx 1.14(1)$.

   Finally we have studied a mixture of the Minimal Wealth and Maximal Wealth models. At every bipartite 
trade the first agent with minimal wealth is selected with probability $p$ or with maximal wealth 
with probability $1-p$. The second agent is selected with uniform probability from the neighbors of the
first agent. A snapshot of the individual wealth for $p=1/2$ at the stationary state for different agents 
has been shown in Fig. 14(b). Here the wealth values are restriced within a `wealth-band' with 
sharp cut-offs at a high and a low end. Consequently the shape of the wealth distribution $P(w,p)$ 
has double peaks for all $N$ and we plot the distribution in Fig. 16(a) for $N=2^9$. Portion of the distribution 
between the peaks fits excellent to elevated Gaussian distributions with different parameter values for different values of $p$ in the 
range between 1/2 and 1. On the two sides of this region the distribution decays to zero very fast. 
The figure shows the plot of $P(w,p)$ vs. $w$ for $p$ = 0.02, 0.08, 0.2, 0.4, 0.6 and 0.9. In Fig. 16(b)
we plot the ratio $r(p)$ of the heights of right peak and the left peak with the probability $p$.

\section {4. Conclusion}

      Social inequality in terms of economic strengths is ubiquitous for the people of all countries. Perhaps this inequality acts as the major 
   driving force behind the advancement of society. Consequently the mechanism that establishes this inequality in a society is an important issue 
   and attracted the attention of researchers over the last century. Here we have studied a modified version of the conservative self-organized 
   extremal model introduced by Pianegonda et. al. which is motivated by the wealth distribution in a society. 
   In this model the entire wealth of the society is strictly conserved. It evolves by a trade dynamics that takes the 
   society from equality (or any other initial wealth distribution) to a stationary state where strong social inequality is present. The dynamics is an 
   infinite sequence of stochastic bipartite tradings where one of the agents has the globally minimal value of wealth, the other one being selected 
   randomly from the neighbors of the first agent. Our numerical study reveals that this model is one of the simplest models of Self-organized 
   Criticality where the
   stationary state is non-ergodic. This model is very similar to the self-organized critical Bak-Sneppen model for the ecological evolution of 
   interacting species. Using numerical simulation we have estimated a number of critical exponents for this model on an $1d$ regular lattice, 
   $2d$ square lattice, the Barab\'asi - Albert scale-free graph and on the $N$-clique graph. We present evidences which suggest 
   that this model does not belong to the universality class of either the Bak-Sneppen model or the Manna model of Self-organized Criticality. 
   This model belongs to a new universality class perhaps because of strict conservation of wealth is maintained in its dynamical rules.

\vskip 0.3 cm
      G. Mukherjee thankfully acknowledges the associateship in S. N. Bose National Centre for Basic Sciences, Kolkata under the Extended Visitor and Linkage 
   programme. Comments from B. K. Chakrabarti, A. Chatterjee, D. Dhar, J. R. Iglesias, J. Kertesz, P. K. Mohanty and P. Pradhan are thankfully acknowledged.

\vskip 0.3 cm
\leftline {manna@bose.res.in}

\end {document}